\documentstyle[12pt,psfig]{article}
\def\be{\begin{eqnarray}}
\def\ee{\end{eqnarray}}

\begin{document}
\thispagestyle{empty}
\begin{flushright}
{\bf TTP96-26\footnote{The complete postscript file of this
preprint, including figures, is available via anonymous ftp at
ttpux2.physik.uni-karlsruhe.de (129.13.102.139) as
/ttp96-26/ttp96-26.ps or via www at
http://ttpux2.physik.uni-karlsruhe.de/cgi-bin/preprints/ 
Report-no: TTP96-26}}\\ 
{hep-ph/9607259}\\
{\bf July 1996}\\
\end{flushright}
\vspace*{20mm}
\begin{center}
  \begin{large}
Higher-Order Hadronic Contributions to the Anomalous Magnetic Moment of Leptons\\
  \end{large}
  \vspace{0.5cm}
  \begin{large}
Bernd Krause
 \\
  \end{large}
  \vspace{0.1cm}
Institut f\"ur Theoretische Teilchenphysik, \\
Universit\"at Karlsruhe,
D-76128 Karlsruhe, Germany \\[4mm]

\vspace{40mm}
  {\bf Abstract}\\
\vspace{0.3cm}
\noindent
\begin{minipage}{15.0cm}
Higher-order hadronic contributions to the anomalous magnetic moment
of the electron, muon, and $\tau$ lepton are considered in 
detail. As a main result we find a reduction by $-11 \times 10^{-11}$
for the g-2 of the muon as compared to previous calculations.
Analytical expressions for the kernel functions of higher-order
hadronic effects are presented. We employ the method of asymptotic
expansions to calculate kernel functions analytically 
in terms of a series expansion in the lepton mass.
\end{minipage}
\end{center}
\newpage
\setcounter{page}{1}

\noindent
The investigation of the anomalous magnetic moment of the muon has
received considerable interest recently due to the upcoming BNL
experiment \cite{Hughes92}. The main goal of this experiment is the 
confirmation of electroweak loop effects as predicted by the Standard
Model. At the same time, deviations may be an indication of physics
beyond the Standard Model. It is therefore necessary to fix the
Standard Model prediction as precisely as possible.  

In this context we note that not only electroweak loop effects but
also higher-order hadronic contributions come into the range of
experimental detection. The designated experimental uncertainty of
$\pm 40\cdot 10^{-11}$ in the Brookhaven experiment allows for an
establishment of higher-order hadronic effects by at least 4 standard
deviations. For the electron the situation is similar: from
improvements of the Penning-trap technique, experimentalists expect a
reduction of the present experimental uncertainty from $\Delta a_e =\pm
4.0\cdot 10^{-12}$ to $\Delta a_e =\pm 0.6\cdot 10^{-12}$
\cite{kin96}, in which case hadronic contributions to $a_e=(g_e-2)/2$
would be a $2\sigma$ effect.

\noindent
Present experimental limits for the $\tau$ lepton are of the order of
$|\Delta a_\tau |\approx 6\cdot 10^{-3}$ \cite{MP94,WM94}, nearly
three orders of magnitude below the expected hadronic effects.
\newline

The purpose of the present work is to calculate 
higher-order hadronic diagrams for the electron, muon, and $\tau$
lepton. A new method for the computation of kernel functions is
presented. Combined with previous work we now provide a complete set of
analytic kernel functions that allow a numerically reliable evaluation
of higher-order hadronic contributions to the g-2 of leptons.

This work is based on a recent analysis of experimental 
data by Eidelman and Jegerlehner \cite{Jeg95}.
We follow this reference in the treatment of integration over hadronic
data. In particular, we also use the trapezoidal rule for the
integration over the $\rho$ resonance. Higher resonances are included
using the narrow width approximation.
The set of kernel functions presented here can, however, also be used
for other parametrizations of hadronic data.

The relevant Feynman diagrams are depicted in fig. (2). 
As in the case of the leading order hadronic correction (fig. 1),
these diagrams cannot be computed in QCD perturbation theory. Instead,
experimental data on the branching ratio $R(s)$ for
$e^+ e^- \rightarrow hadrons$ can be used to compute the diagrams of 
figs. (1) and (2a--c). This is not possible for the so-called
light-by-light-diagram (fig. 3). The latter has been computed 
recently for the muon by two
groups using the extended Nambu-Jona-Lasinio model 
\cite{haya95,bijn95}. It will not be discussed 
in the present paper. 
\newline

The paper is organized as follows: 
In the first section we discuss the calculation of kernel functions
and collect relevant formulae. In the following sections we apply these
results to the computation of higher-order hadronic
contributions to the muon, electron, and tau, respectively. Whenever
possible, different methods (analytical and numerical) have been used
as a cross-check.

\subsection*{Kernel Functions}

The common feature of all hadronic contributions discussed in this
paper will be that they are self-energy insertions in the
photon propagator. We introduce the polarization function
$\Pi_{\mu\nu}$ for the hadronic insertion with
\be
\Pi_{\mu\nu}\left(q\right)&=& i\left(q_\mu
q_\nu-g_{\mu\nu}q^2\right)\Pi(q^2)\\
\Pi(q^2)&=&-{q^2\over\pi}\int_{4m_\pi^2}^\infty {ds\over s}{{\rm
Im}\Pi(s)\over q^2-s} \; .
\ee
The substitution rule from a photon propagator to a
renormalized propagator with hadronic insertion follows:
\be
-i{g_{\mu\nu}\over q^2}\rightarrow {1\over\pi}\int_{4m_\pi^2}^\infty
ds {{\rm Im}\Pi(s)\over s}\left(-i{g_{\mu\nu}\over q^2-s}\right)\; .
\ee

The calculation of higher-order kernel functions reduces to 
the computation of two-loop QED diagrams where one photon is massless and
the other photon has the effective mass $\sqrt{s}$. The 
$s$--integration, which has been shifted to the very end,
is performed over experimental data using 
\be
R(s)&=&{\sigma (e^+e^-\to hadrons)\over \sigma_0(e^+e^-\to\mu^+\mu^-)}
= 12\pi\;{\rm Im}\Pi(s) \nonumber
\ee
and the Born cross section $\sigma_0(e^+e^-\to\mu^+\mu^-)=
{4\pi\alpha^2\over 3 s}$.
\noindent
The leading-order hadronic contribution (fig. 1) to the anomalous
magnetic moment of a lepton $a_l$ is given by
\be
a^{(1)}_l={1\over 3}\left({\alpha\over\pi}\right)^2
\int_{4m_\pi^2}^\infty ds{R(s)K^{(1)}(s)\over s}
\ee
with
\be
K^{(1)}(s)=\int_0^1 dx{x^2(1-x)\over x^2+(1-x){s\over m^2}}\;\; ,
\ee
where $m$ denotes the mass of the external fermion; explicit
expressions for $K^{(1)}(s)$ can be found e.g. in \cite{Jeg95,BR68}.
After renormalization, the higher-order kernel functions for the 
diagrams in fig. (2a,b) can be defined in a similar way:
\be
a^{(2)}_l={1\over 3}\left({\alpha\over\pi}\right)^3
\int_{4m_\pi^2}^\infty ds{R(s)K^{(2)}(s)\over s} \; .
\label{kern2}
\ee

Once the functions $K^{(2)}(s)$ are known analytically, the
higher-order hadronic contributions to the g-2 of leptons can be
computed on an equal footing with the leading order (e.g. with the
same systematic error).
\newline

\noindent
The present situation concerning $K^{(2)}$ is the following:
For the class of diagrams in fig. (2a), a complete analytical result for
the kernel function has been given in \cite{BR75} (eq.(3.21)).
In that reference an expansion to first order in $m^2/s$ has also 
been provided. 
This expansion has been employed for the calculation 
of hadronic contributions from diagrams (2a). 
In the case of diagrams (2b) and (2c) the integral representations
eq.(\ref{num}, \ref{fig2c}) have been used.
\newline

\noindent
As a new approach we employ the method of asymptotic
expansions \cite{Smi94}. This technique has recently been applied to the
calculation of two-loop electroweak corrections to the g-2 of the
muon \cite{CaKrMa95,CaKrMa96}. The calculation of hadronic kernel functions
for the muon and the electron can be treated in exactly the same way. 

The result is an expansion in
the small parameter $m^2/s$, i.e. the ratio of the muon (or electron) mass
m and the energy $\sqrt{s}$ of the final dispersion integral over
hadronic data. 
This final integration is numerically well-behaved
over the whole region of integration.

For the $\tau$ lepton such an expansion cannot be performed since the
integration over hadronic data begins with $\sqrt{s}=2 m_\pi\ll m_\tau$.
However, it is possible to derive an analytic formula for the diagram 
(2b) which describes the contribution for the $\tau$ with sufficient
accuracy. 
\newline

\noindent
As a first step, we repeat the calculation of the kernel function for
the sum of the 14 diagrams in fig. (2a) using the asymptotic
expansions method and confirm 
the result of \cite{BR75}. The expansion up to fourth order reads
\be
K^{(2a)}(s)&=& 2{m^2\over s}\left\{
 \left[{223\over54}-2\zeta(2)-{23\over36}\ln
{s\over m^2}\right]\right. \label{fig1c}\\
&&+ {m^2\over s}\left[{8785\over 1152}-{37\over8}\zeta(2)
-{367\over216}\ln{s\over m^2}
+{19\over144}\ln^2{s\over m^2}\right] \nonumber \\
&&+ {m^4\over s^2}\left[{13072841\over 432000}-{883\over40}\zeta(2)
-{10079\over3600}\ln{s\over m^2}
+{141\over80}\ln^2{s\over m^2}\right] \nonumber \\
&&+ \left.{m^6\over s^3}\left[{2034703\over 16000}-{3903\over40}\zeta(2)
-{6517\over1800}\ln{s\over m^2}
+{961\over80}\ln^2{s\over m^2}\right]\right\} \; .\nonumber 
\ee

\noindent
We now discuss diagram (2b): In the following, $m$ denotes the mass of
the external 
fermion, $m_f$ the mass of the fermion in the loop.
The result of asymptotic expansions including fifth order is
\be
K^{(2b)}(s) &=& 2{m^2\over s}\left\{   \left(-{1\over 18} 
                  + {1\over 9}\ln{m^2\over m_f^2}\right)\right.\nonumber\\
    &&\hspace{-.8cm} + {m^2\over s}\left(- {55\over 48} +{\pi^2\over 18}
             + {5\over 9}\ln{s\over m_f^2} + {5\over 36}\ln{m^2\over m_f^2}
             - {1\over 6}\ln^2{s\over m_f^2}
             + {1\over 6}\ln^2{m^2\over m_f^2}\right)\nonumber\\
    &&\hspace{-.8cm} + {m^4\over s^2}\left(- {11299\over 1800} +{\pi^2\over 3}
             + {10\over 3}\ln{s\over m_f^2} -{1\over 10}\ln{m^2\over m_f^2}
             - \ln^2{s\over m_f^2}
             + \ln^2{m^2\over m_f^2}\right)\nonumber\\
    &&\hspace{-.8cm} - {m^6\over s^3 } \left(
              {6419\over 225}
             - {14\over 9}\pi^2
             + {76\over 45}\ln{m^2\over m_f^2}
             - {14\over 3}\ln^2{m^2\over m_f^2}
             - {140\over 9}\ln{s\over m_f^2}
             + {14\over 3}\ln^2{s\over m_f^2}\right)\nonumber\\
    &&\hspace{-.8cm}\left. - {m^{8}\over s^4 } \left(
              {53350\over 441}
             - {20\over 3}\pi^2
             + {592\over 63}\ln{m^2\over m_f^2}
             - 20\ln^2{m^2\over m_f^2}
             - {200\over 3}\ln{s\over m_f^2}
             + 20\ln^2{s\over m_f^2}\right)\right\}\nonumber\\
    &&\hspace{-.8cm} +2{m_f^2\over m^2}\left[
               {m^2\over s } - {2\over 3}{m^4\over s^2}
              - {m^6\over s^3 } \left(
              - 2\ln{s\over m^2}
              + {25\over 6}\right)
              - {m^8\over s^4 } \left(
              - 12\ln{s\over m^2}
              + {97\over 5}\right)\right.\nonumber\\
    &&\hspace{-.8cm}  \left.\hspace{1cm} - {m^{10}\over s^5 } \left(
              - 56\ln{s\over m^2}
              + {416\over 5}\right)
               \right]\; .
\label{fig1b}
\ee
\noindent
$K^{(2a)}$ and $K^{(2b)}$ can be inserted directly in eq.(\ref{kern2}).

\noindent
For diagrams (2b) there are other representations (numerical and
analytical) which may be used to
check the result from asymptotic expansions.

There are two possible starting points for numerical work:
\be
a^{(2b)}={\alpha\over2\pi^4}\int_{4
m_\pi^2}^{\infty}ds\,\sigma_H(s)\int_0^1dx\frac{x^2
(1-x)}{x^2+(1-x){s\over m^2_\mu}}\cdot\Pi({m_f^2\over m^2})\; ,
\label{num}
\ee
with either \cite{r4}
\be
\Pi\left({m_f^2\over m^2}\right)&=&-{8\over 9}+{\beta^2\over 3}-\left({1\over
2}-{\beta^2\over 6}\right)\beta\ln \frac{\beta-1}{\beta+1}
\label{num1}\nonumber\\
\beta &=& \sqrt{1+4{1-x\over x^2}{m_f^2\over m^2}}
\ee
or \cite{CNPR76}
\be
\Pi\left({m_f^2\over m^2}\right)&=&2 \int_0^1 dy \cdot
y(1-y)\ln\left[1-\left({-x^2\over 1-x}\right){m^2\over
m_f^2}\;y(1-y)\right] \; .\label{num2}
\ee

\noindent
Finally, we derive an analytical expression for the kernel function,
starting from eq.(\ref{num1}). 
Inspection of eq.(\ref{fig1b}) shows that, for light fermions in the loop,
the terms suppressed with ${m_f^2\over m^2}$ are
numerically negligible. We therefore expand $\beta$ in eq.(\ref{num1})
in ${m_f^2\over m^2}$ and retain only unsuppressed terms. After
that, the $x$ integration can be performed and we obtain

\be
K^{(2b)}_{analytical}(s)&=&
-\left({10\over 9}+{2\over3}\ln{m_f^2\over m^2}\right)
\times \label{an2b}\\
&&\hspace{-2.5cm}\times\left\{{1\over2}-(x_1+x_2)
+{1\over x_1-x_2}\left[
x_1^2(x_1-1)\ln\left({-x_1\over 1-x_1}\right)
-x_2^2(x_2-1)\ln\left({-x_2\over 1-x_2}\right)\right]\right\}
\nonumber\\
&&\hspace{-1.5cm} -{5\over6}+{2\over3}(x_1+x_2)+{2\over 3(x_1-x_2)}
\left\{
x_1^2(1-x_1)\left[{\rm Li}_2\left({1\over x_1}\right)
-{1\over2}\ln^2\left({-x_1\over 1-x_1}\right)\right] 
\right.\nonumber\\
&&\hspace{4.0cm}\left.-x_2^2(1-x_2)\left[{\rm Li}_2\left({1\over x_2}\right)
-{1\over2}\ln^2\left({-x_2\over 1-x_2}\right)\right]\right\}\; ,\nonumber
\ee
\noindent
with $b={s\over m^2}$ and $x_{1,2}$ and the dilogarithm defined as follows:
\be
x_{1,2}&=&{1\over2}(b\pm\sqrt{b^2-4b}) \nonumber\\
{\rm Li}_2(t) &=& -\int_0^t dx {\ln(1-x)\over x} \; .\nonumber
\ee

\noindent
This expression is exact up to terms of order ${\cal O}({m_f^2\over
m^2})$. In particular, we still have the full $s$ dependence.
Upon expansion of $K^{(2b)}_{analytical}(s)$ in ${m^2\over s}$ we recover the
unsuppressed part of eq.(\ref{fig1b}).
\newline

\noindent
For the diagram (2c) with two hadronic self energy insertions on the photon
propagator the representation is \cite{CNPR76}
\be
a^{(2c)}&=&{1\over16\pi^5\alpha}
\int_{4 m_\pi^2}^{\infty}ds\;\sigma_H(s)
\int_{4 m_\pi^2}^{\infty}ds'\sigma_H(s')
\int_0^1dx\frac{x^4(1-x)}
{[x^2+(1-x){s\over m^2}][x^2+(1-x){s'\over m^2}]}
 \nonumber\\
&=& {\alpha^3\over9\pi^3}
\int_{4 m_\pi^2}^{\infty}dsds'{R(s)R(s')K^{(2c)}(s,s')\over s\cdot s'}
\; .\label{fig2c}
\ee
\noindent
An analytical expression for $K^{(2c)}(s,s')$ is easily found; 
setting $b={s\over m^2}$ and
$c={s'\over m^2}$ we obtain for $b\neq c$:
\be
K^{(2c)}(s,s')&=&
{1\over 2} - b - c - {{\left( 2 - b \right) \,{b^2}\,\ln (b)}\over 
    {2\,\left( b - c \right) }} - 
  {{{b^2}\,\left( 2 - 4\,b + {b^2} \right) \,
      \ln ({{b + {\sqrt{- \left( 4 - b \right) \,b  }}}\over 
         {b - {\sqrt{- \left( 4 - b \right) \,b  }}}})}\over 
    {2\,\left( b - c \right)
     {\sqrt{- \left( 4 - b \right) \,b  }}\,
       }} \nonumber \\
&& - 
  {{\left( -2 + c \right) \,{c^2}\,\ln (c)}\over 
    {2\,\left( b - c \right) }} + 
  {{{c^2}\,\left( 2 - 4\,c + {c^2} \right) \,
      \ln ({{c + {\sqrt{- \left( 4 - c \right) \,c }}}\over 
         {c - {\sqrt{- \left( 4 - c \right) \,c }}}})}\over 
    {2\,\left( b - c \right) \,{\sqrt{- \left( 4 - c \right)  \,c
          }}}} \; ,
\label{form2c1}
\ee
the corresponding formula for $b=c$ is
\be
K^{(2c)}(s,s')&=&
{1\over 2} - 2\,c 
+{c\over2}\left( -2 + c - 4\,\ln (c) + 3\,c\,\ln (c) \right)
  \label{form2c2}\\
&& + {{{c\,\left( -2 + 4\,c - {c^2} \right) }
\over {2(-4 + c)}} +    {{c\,\left( 12 - 42\,c + 22\,{c^2} - 3\,{c^3} \right) \,
          \ln ({{c + {\sqrt{\left( -4 + c \right) \,c}}}\over 
             {c - {\sqrt{\left( -4 + c \right) \,c}}}})}\over 
        {2\left( -4 + c \right) \,{\sqrt{\left( -4 + c \right) \,c}}}}}
\nonumber
\ee

\subsection*{Muon}

We take the value for the leading-order hadronic contribution to $a_\mu$ 
from \cite{Jeg96} (see \cite{AY95,WB95} for other recent analyses)
\be
a_\mu^{(1)} = 7023.5(58.5)(140.9)\cdot10^{-11} \;.
\ee
Here and in the following the first bracket gives the statistical, the
second the systematic error. For the higher-order effects we use the
combined statistical and systematic error of the leading order as an
error estimate.

Higher-order hadronic contributions to the g-2 of the muon
were considered for the first
time in 1976 \cite{CNPR76} and reevaluated in 1984 
\cite{kno} with new input from experimental data and a new
calculation of the light-by-light diagram. 

We start with a discussion of the contribution from fig.(2a),
employing the expansion eq.(\ref{fig1c}).
If we take only the first line in this formula (this corresponds to 
eq. (3.25) in \cite{BR75}) we obtain
\be
a^{(2a)}_\mu({\rm first\; order})=-200(4)\cdot 10^{-11}\; ,
\label{erg1a}
\ee
which is very close to the value of $-199\cdot 10^{-11}$ in \cite{kno}.
Including the higher-order terms shifts the number to
\be
a^{(2a)}_\mu=-211(5)\cdot 10^{-11} \; .
\label{erg2a}
\ee

We checked the integration over the expanded kernel function against
the full formula of ref.\cite{BR75}. The deviation is below 0.3\% if 
at least four orders
in the expansion are included. It should be noted, however, that the
numerical evaluation of eq.(3.21) in \cite{BR75} for large values of
$s$ (i.e. in a region where the expansion converges rapidly) is less
reliable than the expansion itself.
\newline

\noindent
We now turn to the contribution from the diagrams of fig. (2b). We restrict
ourselves to the case of an electron in the loop. The contribution of
the $\tau$ loop is suppressed by $m_\mu^2\over m_\tau^2$. Using the
kernel function of eq.(\ref{fig1b}) we obtain in agreement with \cite{kno}
\be
a^{(2b)}_\mu= 107(2)\cdot 10^{-11} \; .
\label{erg2b}
\ee

Again we notice that at least four or five terms in the expansion must
be included in order to obtain a stable result.
The number decreases from $126\cdot 10^{-11}$ in first order to
the value of eq.(\ref{erg2b})
in fifth order (between 4th and 5th order the value changes by 0.2\%).
The part of eq.(\ref{fig1b}) suppressed with ${m_e^2\over m_\mu^2}$
turns out to be numerically completely negligible.

The result of eq.(\ref{erg2b}) has been checked using different 
numerical integration
methods and the analytical representation of eq.(\ref{an2b}).
Whereas the analytical result coincides with
the expansion, care has to be taken in the numerical integrations.

Numerically, the representation eq.(\ref{num1}) is much less appropriate
than eq.(\ref{num2}). The latter is a flat function and the double
numerical integration over $x$ and $y$ yields the result
in eq.(\ref{erg2b}). The representation eq.(\ref{num1}) is a peaked
function and although for fixed $s$ the error is at worst of the
order of 1\% it adds up systematically in the $s$ integration and
reaches 10\% in the final result.
\newline

To conclude the discussion for the muon, we give the value we obtain
for diagram (2c) with two hadronic bubbles on the photon
propagator. Using eq.(\ref{fig2c}) we obtain
\be
a^{(2c)}_\mu= 2.7(0.1)\cdot 10^{-11} \; .
\label{erg2c}
\ee
This number can be reproduced using the analytical expressions
eq.(\ref{form2c1},\ref{form2c2}) with quadruple precision Fortran. 

We find a total shift in the higher-order hadronic contributions of
$a_\mu$ from \cite{kno}
\be
a_\mu^{(2a+2b+2c)}=-90(5)\cdot 10^{-11}
\ee
to
\be
a_\mu^{(2a+2b+2c)}=-101(6)\cdot 10^{-11} \; .
\ee
The difference $(-11)\cdot 10^{-11}$ is somewhat smaller but still of
the same order of magnitude as the 
experimental uncertainty in the BNL experiment, or approximately 25\%
of the two-loop electroweak contributions. It should be emphasized
that the reason for the shift doesn't reside predominantly in the
input of new experimental data;
rather, the high precision of the Brookhaven experiment requires the
use of the full kernel functions.

\subsection*{Electron}

Higher-order hadronic contributions to the g-2 of the electron have
only recently \cite{kin96} been considered for the first time. Some
features are special to the electron (as opposed to the muon case)
and will be discussed in the following.

\noindent
For the electron the calculation of diagram (2a) is very similar to
the muon case. It is sufficient to take the first term in the
expansion of eq.(\ref{fig1c}).
\be
K^{(2a)}(s) = - 2{m_e^2\over s}\left( {23\over36} \ln{s\over m_e^2} 
       +{\pi^2\over3} -{223\over54}\right)\; .
\ee
Here, the contribution from the large logarithm
dominates in the whole region of integration.
We obtain the value
\be
a_e^{(2a)} = -2.25(0.05)\cdot10^{-13} \; .
\ee

\noindent
The diagrams of fig. (2b) with an external electron yield a numerically
negligible contribution. The first observation is that the diagrams
with an electron self-energy on the photon propagator are already
contained in the set of diagrams (2a). Diagrams with a muon or a
$\tau$ lepton on the photon propagator are suppressed by
additional powers of $m_e^2\over m_\mu^2$ or $m_e^2\over m_\tau^2$ 
respectively, and hence can be omitted.
This is a special case of the general statement that in diagrams of
the type (2b) large logarithms occur if the mass of the external
fermion is bigger than the mass of the fermion in the loop (see, for
instance, the leading term $ {1\over 9}\ln{m_\mu^2\over m_e^2}$ in
eq.(\ref{fig1b})); in the opposite case there is a mass suppression.
The double hadronic bubble diagram (2c) is of the
order $a_e^{(2c)} \approx 1.2\cdot10^{-19}$ for the same reason.
\newline
\noindent
For an estimate of the total higher-order hadronic contribution to
$a_e$ including light-by-light scattering we refer to ref. \cite{kin96}:
\be
a_e^{(2a)}+a_e^{(3)} = ( -2.25 -0.122)\cdot10^{-13} =
-0.24\cdot10^{-12}\; ,
\ee
where the light-by-light value has been estimated by rescaling the 
corresponding number for the muon.

The higher-order corrections to $a_e^{\rm had}({\rm leading})$
\cite{Jeg96} 
\be
a_e^{(1)} = 1.8847(0.0165)(0.0375)\cdot10^{-12} 
\ee
reduce the total hadronic contribution by 13\% ; the final number is
\be
a_e^{\rm had}({\rm total}) = 1.645\cdot10^{-12} \;.
\ee

\subsection*{$\tau$ lepton}

The leading-order hadronic effect is \cite{Jeg96}
\be
a_\tau^{(1)} = 338.30\;(1.97)(9.12)\cdot10^{-8} \; .
\ee
In \cite{SL91} an estimate of the higher-order hadronic contributions
to $a_\tau$ was given
\be
a_\tau^{had}({\rm higher\; orders}) = -1.2\;(2)\cdot10^{-7} \; .
\ee

The number can be reproduced by rescaling the muon results of
\cite{kno} with the factor ${m_\tau^2\over m_\mu^2}$ and taking the 1991
value for $m_\tau$: 
\be
{m_\tau^2\over m_\mu^2}
\left(a_\tau^{(2a+2b+2c)}+a_\tau^{light-by-light}\right)=
(-90+49)\cdot 10^{-11}\cdot\left({1784\over 106}\right)^2
= -1.2\cdot 10^{-7} \; .\nonumber
\ee
It will be shown in the following that, as in the case of the
electron, simple rescaling of the muon results doesn't give a correct
estimate. 

\noindent
For the $\tau$ lepton it is no longer possible to use expansions in
${m_\tau^2\over s}$ over the whole region of integration because the
integration starts with $s=4 m_\pi^2$. The diagrams of fig. (2a) can
still be calculated with the general kernel function of \cite{BR75}.
We obtain the value
\be
a_\tau^{(2a)} = -3.18(0.08)\cdot10^{-8} \; .
\ee

In the case of the diagrams of fig. (2b) we have to sum over the 
contributions of the electron and the muon in the loop.
Here, we apply two different methods to compute the anomaly: a
two-dimensional numerical integration yields the results
\be
a_\tau^{(2b)}({\rm electron}) &=& 7.48(0.2)\cdot10^{-8} \\
a_\tau^{(2b)}({\rm muon}) &=& 1.91(0.05)\cdot10^{-8} \; .
\label{veg2b}
\ee
\noindent
Using the analytic formula eq.(\ref{an2b}) we
obtain the following values:
\be
a_\tau^{(2b)}({\rm electron}) &=& 7.48(0.2)\cdot10^{-8} \\
a_\tau^{(2b)}({\rm muon}) &=& 1.88(0.05)\cdot10^{-8} \; .
\ee
As expected, the electron contribution is the same and we have a 2\%
deviation for the muon. This is consistent with the order of magnitude
of the neglected terms, i.e.
\be
{m_\mu^2\over m_\tau^2}\ln{m_\tau^2\over m_\mu^2}\approx 0.02
\ee

For the $\tau$ lepton, diagram (2c) also yields a numerically important
contribution. From eq.(\ref{fig2c}) we get the value 
\be
a_\tau^{(2c)} = 1.4(0.1)\cdot10^{-8} \; .
\ee

The higher-order correction to $a_\tau^{\rm had}({\rm leading})$
enhances the total hadronic contribution by 2\% (without the light-by
light contribution): 
\be
a_\tau^{(2a+2b+2c)} = 7.6(0.2)\cdot10^{-8} \;.
\label{ergtau}
\ee
This should be contrasted with the estimate from simple rescaling:
\be
\left({m_\tau^2\over m_\mu^2}\right)
a_\mu^{(2a+2b+2c)}=
(-101)\cdot 10^{-11}\cdot\left({1777\over 106}\right)^2
= -2.8\cdot 10^{-7} \; .\nonumber
\ee

Keeping this uncertainty in mind, we give an estimate for higher-order
hadronic contributions to $a_\tau$ with
eq.(\ref{ergtau}) and a rescaled light-by-light value from \cite{haya95}:
\be
a_\tau^{had}({\rm higher\; orders}) \approx
\left({m_\tau^2\over m_\mu^2}\right)\cdot(-52\cdot 10^{-11})
+ 7.6\cdot10^{-8} = -7.0\cdot 10^{-8} \; .\nonumber
\ee

\section*{Conclusions}

We presented analytic formulae for the calculation of higher-order
hadronic contributions to the g-2 of leptons and introduced the method
of asymptotic expansions as a convenient tool to calculate the
corresponding kernel functions. We demonstrated that in order to match
the high experimental precision of the upcoming experiments, it is
necessary to include the full kernel functions in theoretical
predictions. The higher-order hadronic effects on the anomalous
magnetic moments of the muon, electron, and $\tau$ lepton are
presented. The remaining uncertainties in the predictions from the
Standard Model reside now in the experimental error in the measurement
of $\sigma (e^+ e^- \to hadrons)$ and the calculation of
light-by-light-scattering effects.

\section*{Acknowledgements}

I would like to thank F. Jegerlehner and S. Eidelman for the
permission to use their set of experimental data.  I am grateful to
A. Czarnecki, T. Kinoshita, J. H. K\"uhn and T. Teubner for useful
discussions and advice, and M. L. Stong for reading of the
manuscript. I would like to thank A. H\"ocker for checking part of the
numerical results and useful comments.
This work was supported by ``Graduiertenkolleg
Elementarteilchenphysik'' at the University of Karlsruhe.

\newpage

\vspace*{1cm}
\noindent
\begin{minipage}{16.cm}
\hspace*{.8cm}
\[
\mbox{
\hspace*{-0mm}
\psfig{figure=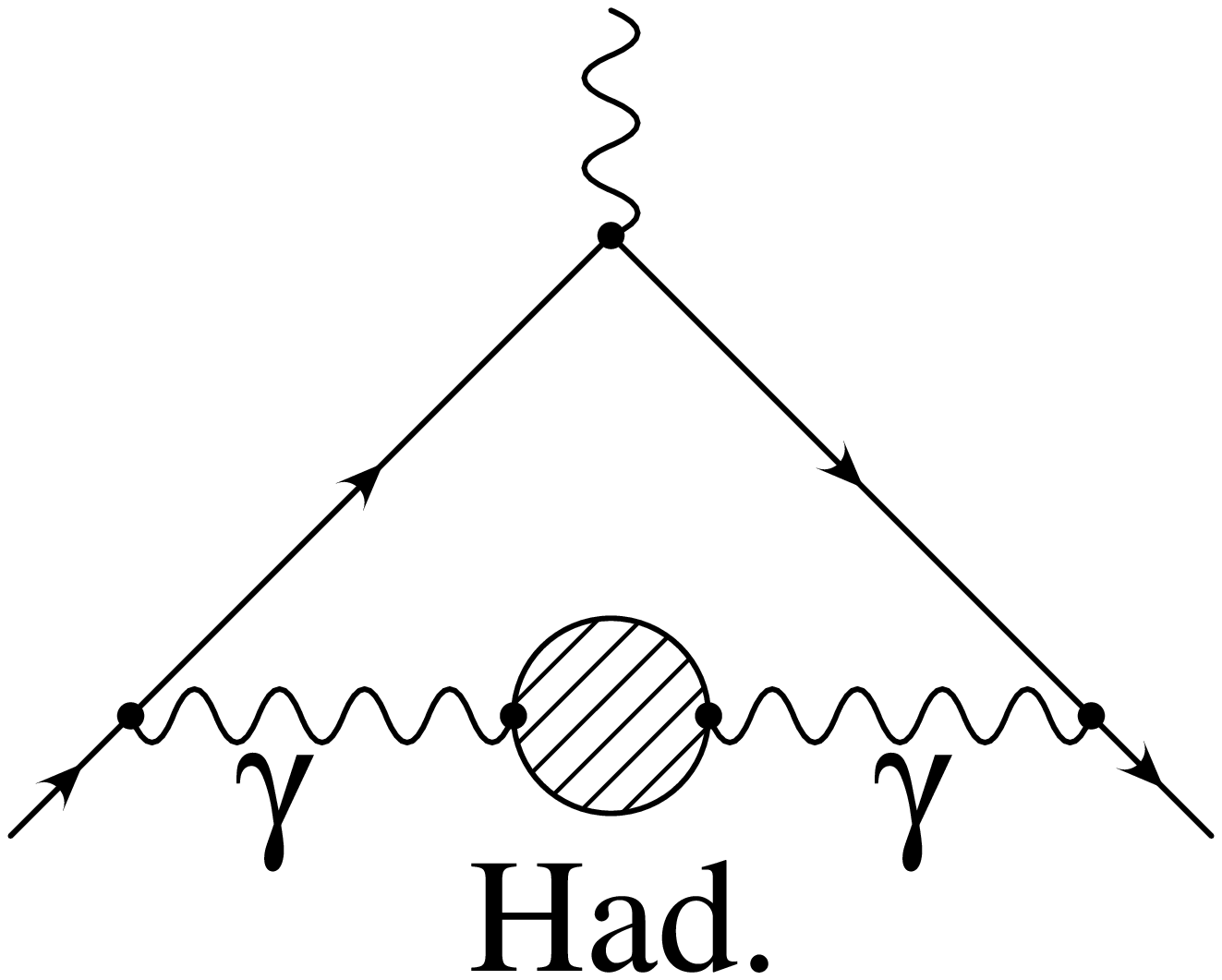,width=25mm,bbllx=210pt,bblly=410pt,bburx=630pt,bbury=550pt} 
}
\]
\end{minipage}
\vspace*{5mm}

\hspace*{1.8cm}
Fig. (1): Leading order hadronic contribution

\vspace*{1cm}
\noindent
\begin{minipage}{16.cm}
\hspace*{.8cm}
\[
\mbox{
\hspace*{-0mm}
\begin{tabular}{ccccc}
\psfig{figure=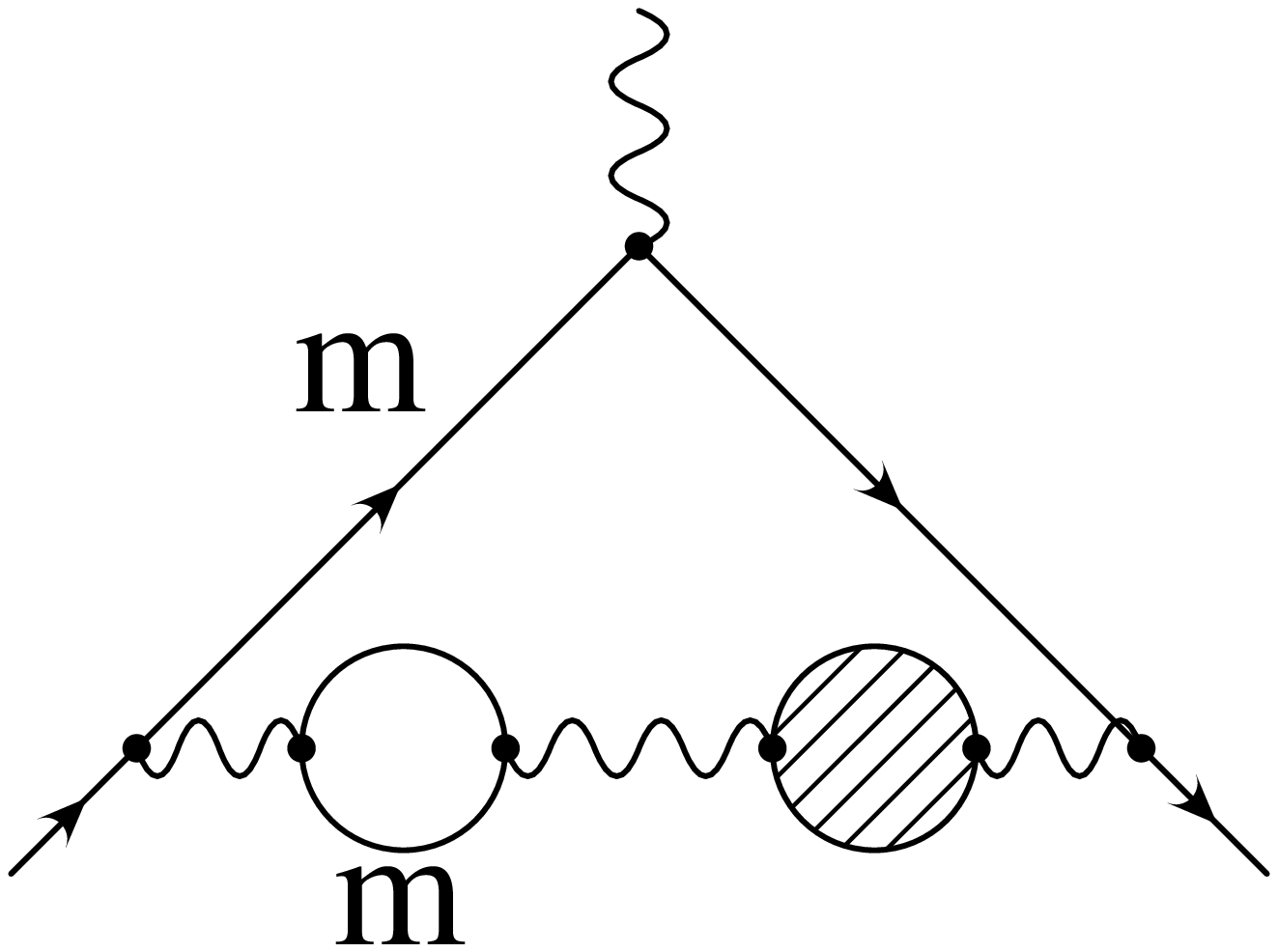,width=25mm,bbllx=210pt,bblly=410pt,bburx=630pt,bbury=550pt} 
&\hspace*{0mm}
\psfig{figure=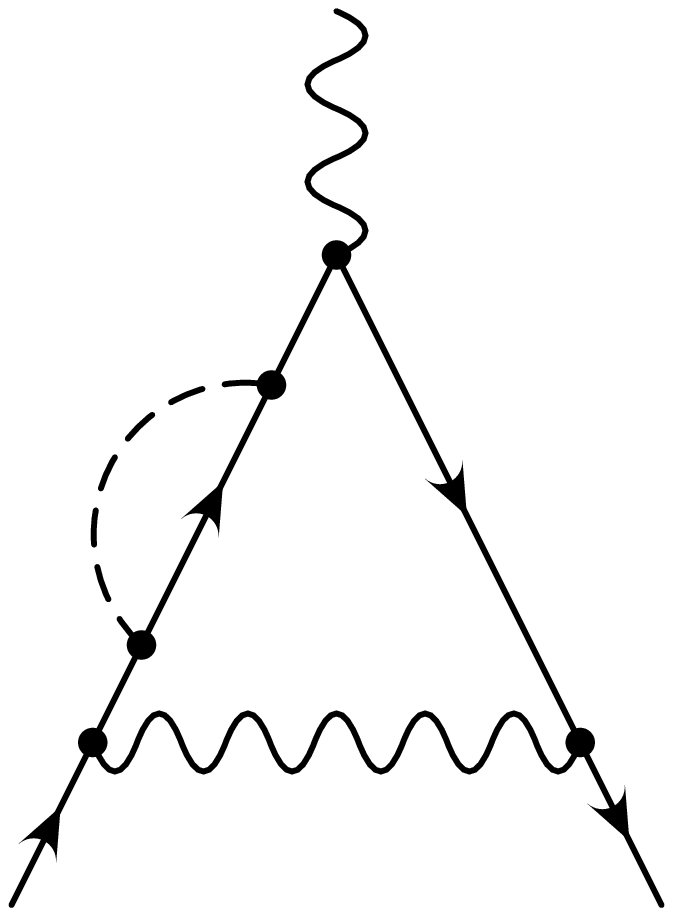,width=25mm,bbllx=210pt,bblly=410pt,bburx=630pt,bbury=550pt}
&\hspace*{0mm}
\psfig{figure=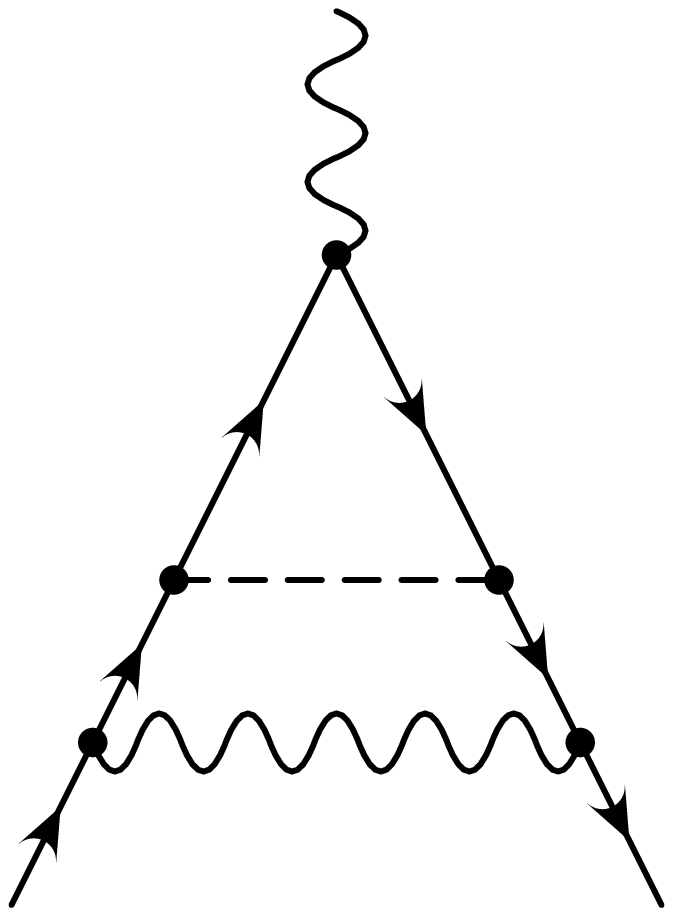,width=25mm,bbllx=210pt,bblly=410pt,bburx=630pt,bbury=550pt}
&\hspace*{0mm}
\psfig{figure=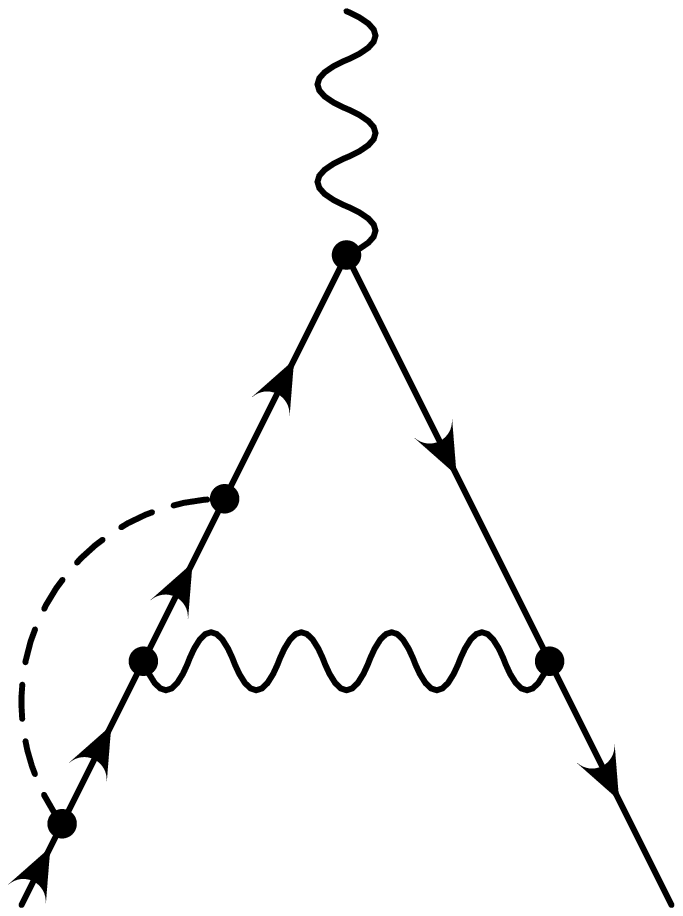,width=25mm,bbllx=210pt,bblly=410pt,bburx=630pt,bbury=550pt} 
&\hspace*{0mm}
\psfig{figure=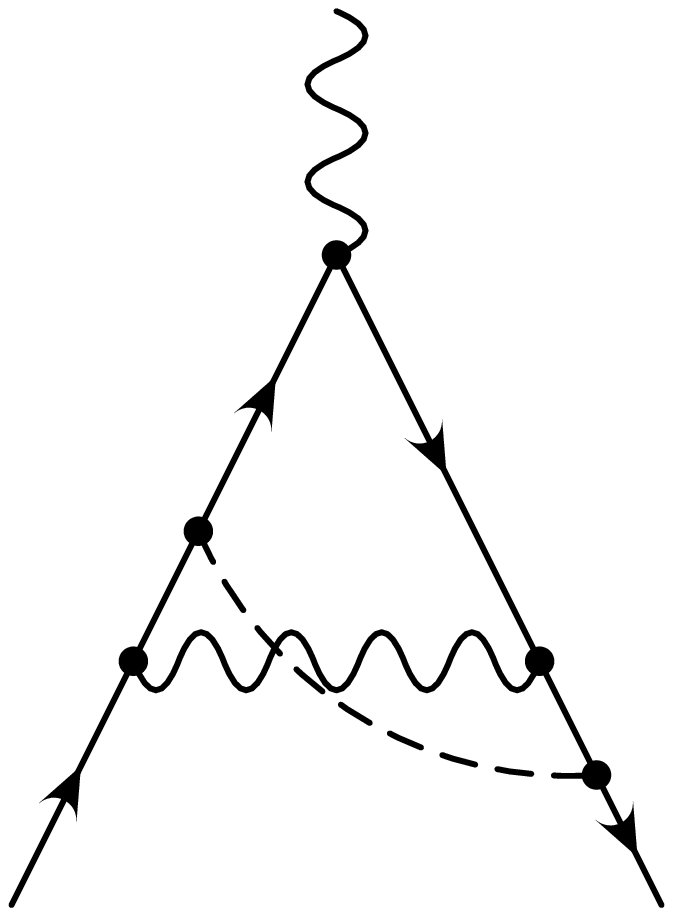,width=25mm,bbllx=210pt,bblly=410pt,bburx=630pt,bbury=550pt}
\end{tabular}
}
\]
\end{minipage}
\vspace*{4mm}

\noindent 
Fig. (2a): Dashed lines indicate a
hadronic insertion on the photon propagator. Mirror counterparts and
diagrams with interchange of massless and ``massive'' photon
propagators have to be included.

\vspace*{1cm}
\noindent
\begin{minipage}{16.cm}
\hspace*{0.8cm}
\[
\mbox{
\hspace*{0mm}
\begin{tabular}{ccc}
\psfig{figure=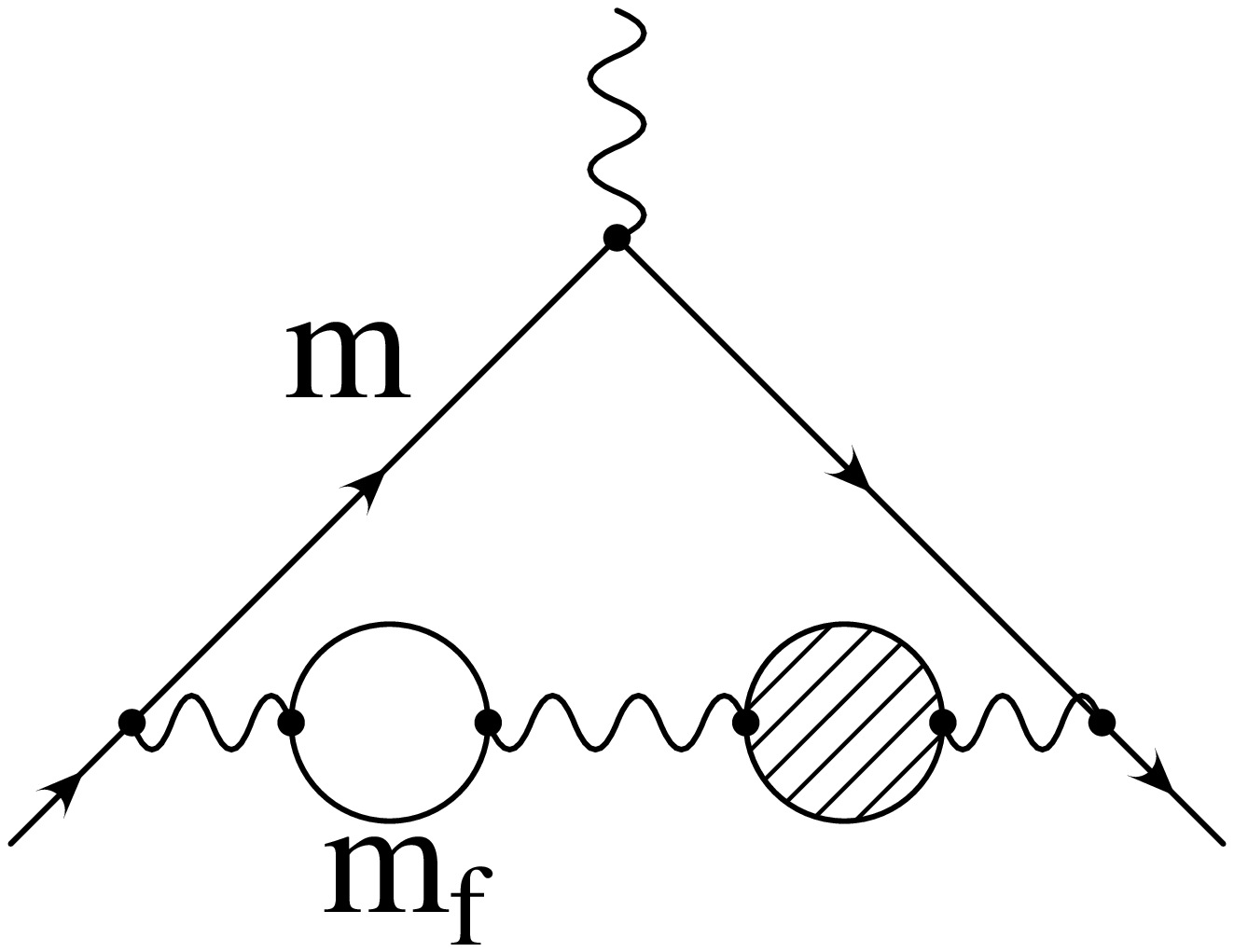,width=28mm,bbllx=110pt,bblly=410pt,bburx=630pt,bbury=550pt} 
&\hspace*{14mm}
\psfig{figure=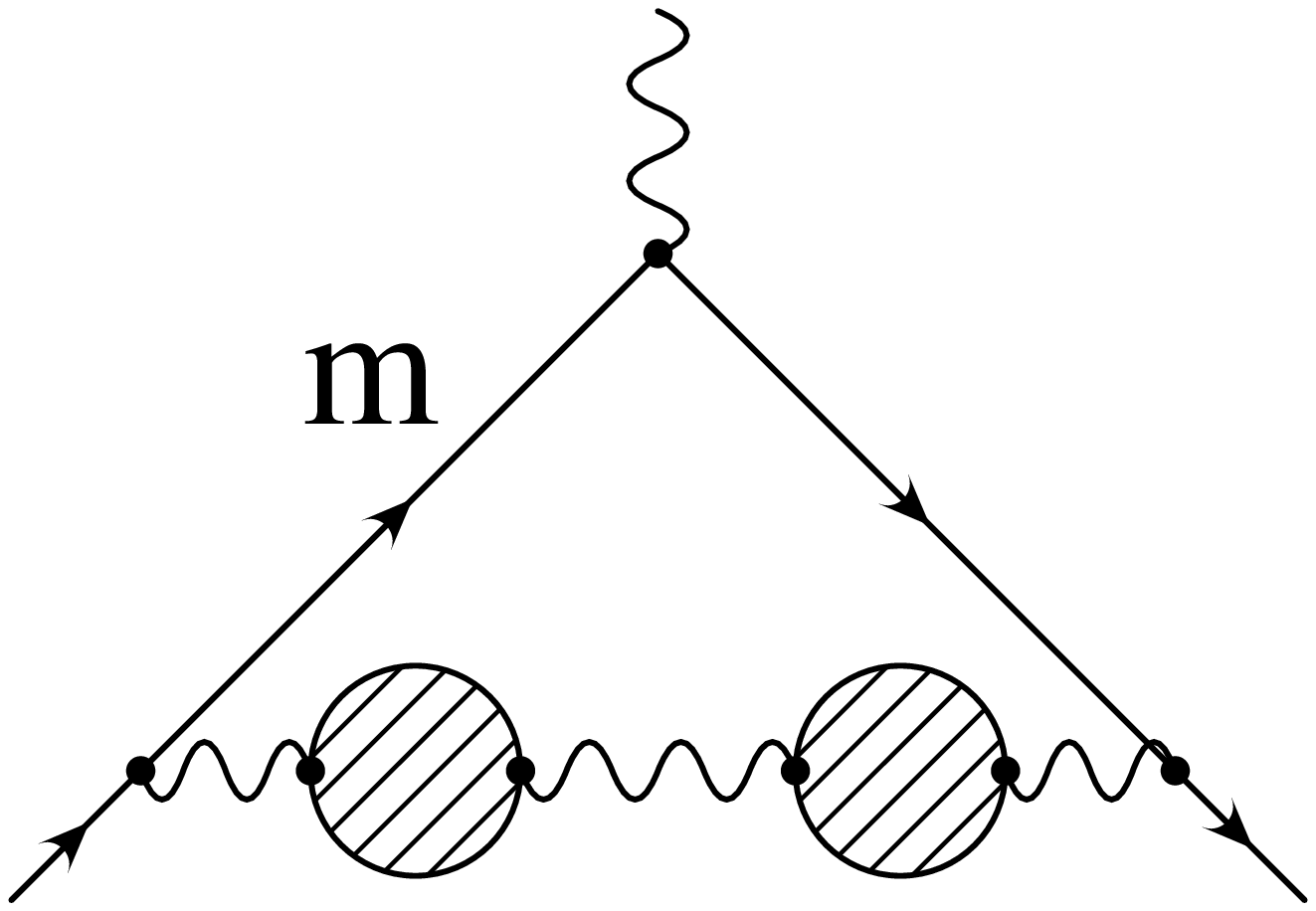,width=28mm,bbllx=110pt,bblly=410pt,bburx=630pt,bbury=550pt}
&\hspace*{14mm}
\psfig{figure=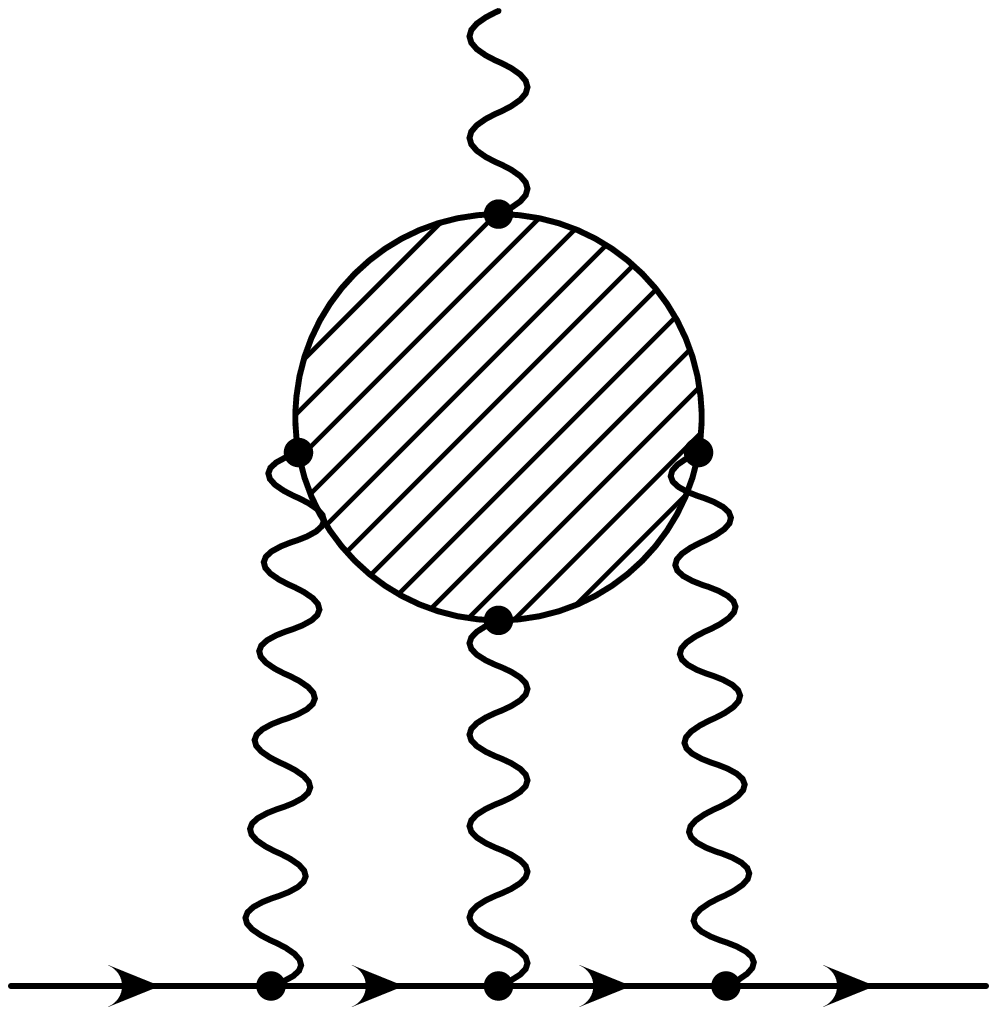,width=28mm,bbllx=110pt,bblly=410pt,bburx=630pt,bbury=550pt}
\\[5mm]
\rule{-12mm}{0mm} Fig. (2b) &\hspace*{6.8mm} \rule{-4mm}{0mm}Fig. (2c) 
&\hspace*{12mm} \rule{-4mm}{0mm}Fig. (3) 
\end{tabular}}
\]
\end{minipage}


\begin{thebibliography}{10}

\bibitem{Hughes92}
V.~W. Hughes,  in {\em Frontiers of High Energy Spin Physics}, edited
by T.~Hasegawa 
  et~al. (Universal Academy Press, Tokyo, 1992), pp.\ 717--722.

\bibitem{kin96}
T. Kinoshita, Cornell preprint CLNS-96-1406 (1996).

\bibitem{MP94}
M. Perl, Nucl. Phys.  {\bf B40} (Proc. Suppl.), 541 (1995).

\bibitem{WM94}
W. Marciano, Nucl. Phys.  {\bf B40} (Proc. Suppl.), 3 (1995).

\bibitem{Jeg95}
S. Eidelman and F. Jegerlehner, Z. Phys. {\bf C67}, 585 (1995).

\bibitem{haya95}
M. Hayakawa, T. Kinoshita, and A.~I. Sanda, Phys. Rev. Lett. {\bf 75},  790
  (1995), and M. Hayakawa, T. Kinoshita, and A.~I. Sanda, DPNU-95-30,
  hep-ph/9601310.

\bibitem{bijn95}
J. Bijnens, E. Pallante, J. Prades, Phys.~Rev.~Lett. {\bf 75}, 1447, (1995);
 Erratum: ibid. {\bf 75}, 3781 (1995), and
 J. Bijnens, E. Pallante, J. Prades, preprint NORDITA-95-75-N-P, 
 hep-ph/9511388.

\bibitem{BR68}
S.J. Brodsky, E. de Rafael, Phys. Rev. {\bf 168}, 1620 (1968).

\bibitem{BR75}
R. Barbieri, E. Remiddi, Nucl. Phys.  {\bf B90},  233  (1975).

\bibitem{Smi94}
For a recent review see
 V.~A.~Smirnov, Mod. Phys. Lett. {\bf A10}, 1485 (1995). 

\bibitem{CaKrMa95}
A. Czarnecki, B. Krause, and W. Marciano, Phys. Rev. {\bf D52}, R2619 (1995).

\bibitem{CaKrMa96}
A. Czarnecki, B. Krause, and W. Marciano, Phys. Rev. Lett. {\bf 76}, 3267 (1996).

\bibitem{r4}
M. Samuel and G. Li, Phys. Rev. {\bf D44}, 3935 (1991);
                     Erratum {\bf D48}, 1879 (1993).

\bibitem{CNPR76}
J. Calmet, S. Narison, M. Perrottet and E. de Rafael, Phys. Lett. {\bf B61}, 283 (1976).

\bibitem{Jeg96}
F. Jegerlehner, in {\em Proceedings of the Zeuthen Workshop on
Elementary Particle Theory: QCD and QED in Higher Orders}, Rheinsberg,
Germany, 21-26 Apr. 1996, preprint DESY-96-121.

\bibitem{AY95}
K. Adel and F. Yndur\'ain, FTUAM 95-2 (1995).

\bibitem{WB95}
W. Worstell and D. Brown, Boston University preprint (1995).

\bibitem{kno}
T. Kinoshita, B. Nizic, and Y. Okamoto, Phys. Rev. {\bf D31}, 2108 (1985).

\bibitem{SL91}
G. Li, R. Mendel and M. Samuel, Phys. Rev. Lett. {\bf 67}, 668 (1991);


\end{thebibliography}
\end{document}